\documentstyle[12pt]{article}

\begin{document}
\renewcommand{\theequation}{\thesection.\arabic{equation}}
\thispagestyle{empty} \vspace*{-1.5cm} {\small hep-th/9909013} \hfill
{\small SU-ITP
99-40}
\\[8mm]

\setlength{\topmargin}{-1.5cm} \setlength{\textheight}{22cm}
\begin{center}
{\large Wilson loops as Precursors}
\\

\vspace{1 cm} {\large Leonard Susskind and Nicolaos Toumbas }\\
Department of Physics, Stanford University, \\ Stanford CA
94305-4060
\\ \vspace{0cm}

\begin{abstract}

There is substantial evidence that string theory on  $AdS_5 \times
S_5$ is a holographic theory in which the number of degrees of
freedom scales as the area of the boundary in Planck units.
Precisely how the theory can describe bulk physics using only
surface degrees of freedom is not well understood. A particularly
paradoxical situation involves an event deep in the interior of
the bulk space.  The event must be recorded in the (Schroedinger
Picture)  state vector of the boundary theory long before a
signal, such as a gravitational wave, can propagate from the event
to the boundary. In a previous paper with Polchinski, we argued
that the ``precursor'' operators which carry information stored in
the wave during the time when it vanishes in a neighborhood of the
boundary are necessarily non--local. In this paper we argue that
the precursors cannot be products of local gauge invariant
operators such as the energy momentum tensor. In fact gauge
theories have a class of intrinsically non--local operators which
cannot be built from local gauge invariant objects. These are the
Wilson loops. We show that the precursors can be identified with
Wilson loops whose spatial size is dictated by the UV--IR
connection.

\end{abstract}
\vspace{1cm} {\it August 1999}
\end{center}

\setcounter{equation}{0}
\section{Introduction}

This paper is concerned with the mechanism by which a holographic
boundary theory can describe bulk physics. As emphasized in
\cite{1}\cite{2}\cite{3} a holographic description entails a vast
reduction of the number of degrees of freedom needed to describe a
region of bulk spacetime. Despite the large amount of
circumstantial evidence for the holographic principle it is still
very mysterious how such a sparse set of degrees of freedom can
describe all local bulk physics. A particular challenge is to
understand how events deep in the interior of the bulk space are
recorded in the instantaneous (Schroedinger Picture) state vector
of the boundary theory long before a signal can propagate from the
event to the boundary \cite{4}.

Let us consider an example. For definiteness we will consider the
3+1 dimensional super Yang Mills description of bulk physics in 5
dimensional AdS space \footnote{The usual $S_5$ factor in the
correspondence plays no role in the present paper and will be
ignored. }\cite{5}\cite{6}\cite{7}.
We will be interested in the limit of large radius of curvature compared to
the
string scale. In this limit stringy excitations are negligible and the
low energy supergravity approximation to bulk physics is reliable.
On the SYM side we must take $N$ large keeping the 't Hooft coupling
constant $g^2N$ fixed and large.

Suppose as in \cite{4} an event takes place at the center
\footnote{Since AdS is a homogeneous space it has no preferred
points. Center here means the origin of cavity coordinates. } of AdS
which radiates a gravitational wave toward the boundary.
No signal including the wave itself can arrive at the boundary until a
certain time elapses.
If the original event is well localized near the center of a very
large AdS space, the original bulk fields will typically be very
spherically symmetric and time independent.
In fact the only bulk field of importance is the time--time component
of the metric whose behavior near the boundary records the presence of
a certain amount of energy in the interior.
On the SYM side this means that the energy momentum tensor is almost
exactly homogeneous and consists of an homogeneous energy density and
the pressure needed to make $\langle{T^{\mu}_{\mu}}\rangle = 0$.
However, this effect is featureless and provides no information about the
profile of the gravitational wave.
In addition, it is vanishingly small in the large $N$ limit since
corrections to
the metric due to the energy of the wave are smaller than the wave
itself by factors of $\sqrt{G_5} \sim N^{-1}$.
We refer the reader in \cite{4} for notations and conventions.
Thus, within a neighborhood of the boundary, all supergravity field
functionals retain their original, vacuum--like expectation
values, at least until light has had a chance to propagate from $r
\sim 0$ to the boundary.
The implication for the  SYM theory is that all expectation values of local
gauge invariant operators  corresponding to the bulk fields, as well
as expectation values of products of such operators, should initially be
identical
to their vacuum values  and contain no information about the propagating wave.

This situation continues until the outgoing wave arrives at the
boundary.
At that time the perturbation on the boundary
becomes nonzero and begins oscillating over the whole 3-sphere.
From the SYM point of view, the energy momentum tensor and its
products suddenly begin to coherently oscillate.
The features of the gravitational wave can then be recovered from
expectation values of the energy momentum tensor and its products.

Thus during the time when the wave vanishes within a
neighborhood of the boundary, the SYM theory is excited to a
non-vacuum-like state which we cannot distinguish from the
vacuum by taking expectation values of local gauge invariant operators or
any of their products\footnote{ We are assuming that local gauge
invariant operators are in
one--to--one correspondence with local observables of the bulk theory
evaluated near the boundary.}.
In \cite{4}, it was argued that the holographic boundary theory must contain
special non-local operators, called precursors, that distinguish such
states and code in detail local bulk information.
The precursors should become increasingly non-local the further the
corresponding bulk process is from the boundary in accordance with the UV--IR
connection \cite{3}.
In those cases in which the boundary theory has a gauge symmetry, the
precursors must also be gauge invariant since they contain
physical information.
In the case of ${\cal{N}} = 4$ SYM theory, this suggests that the
precursors are Wilson loops whose size is dictated by the UV--IR
relation.

We remark that there exists a rich class of generalized, equal--time
Wilson loops being candidates for the precursors.
Apart from conventional spatial Wilson loops, we may consider
spatial Wilson loops with insertions of local gauge covariant operators.
For example, we can consider the operator
\begin{equation}
Tr P F_{\mu\nu}(x_1) F^{\mu\nu}(x_2) W ,
\end{equation}
where $W$ is a Wilson loop passing through the
points $x_1$ and $x_2$  and $P$ denotes path ordering.
Presumably, such operators and their products form a complete set of
observables in the boundary theory.

In \cite{4}, it was shown how a plane gravitational wave can be modeled
by ``squeezed'' states constructed in free field theory.
In particular, it was shown how to account for the oscillating energy
density and the apparent acausality in the behavior of the energy
momentum tensor required by the correspondence.
It was found that this behavior is consistent with bounds required by
general principles of quantum mechanics.
In addition, apart from possible numerical coefficients, the free
field theory model reproduces corrections to the linearized
solution induced by non-linear terms in Einstein's equations involving
the energy density of the wave.

In this note we model bulk waves with ``squeezed states'' constructed
in the interacting SYM theory.
We compute expectation values of local gauge invariant operators
in the ``squeezed states'' and match them with the boundary data of
the wave.
We show that expectation values of products of local gauge
invariant operators contain no additional information about the
profile of the wave in agreement with bulk causality.
Our computations are done in the 't Hooft limit keeping only the
leading terms in the $1/N$ expansion.
Finally, using the corespondence, we calculate expectation values of Wilson
loops in the ``squeezed states'' and show how they carry non-trivial
information if their size is as dictated by the UV--IR
connection.
We discuss the implications of our results for holography at the end.

Before concluding the introduction we will review some facts and
conventions about the AdS--CFT correspondence. The metric of AdS in
cavity coordinates is
\begin{eqnarray}
ds^2 &=& R^2\left[{(1+r^2)^2 \over (1-r^2)^2}dt^2 - {4 \over
(1-r^2)^2}(dr^2 +r^2 d\Omega^2) \right]
\cr &=& R^2 dS^2,
\end{eqnarray}
where the coordinates and $dS^2$ are dimensionless and $d\Omega^2$
is the metric of the unit 3-sphere. The center of AdS means
the point $r=0$. Near a point of the boundary at $r=1$ the metric
has the form
\begin{equation}
ds^2=R^2\left[{1\over z^2}(dt^2 -dz^2 - dx^i dx^i) \right]
\end{equation}
where $z=1-r$ and $x^1,x^2, x^3$ replace the coordinates of the
3-sphere.
For our purposes the metric (1.3) is to be regarded as a local
approximation to the cavity metric.  It is true, but irrelevant to
our purposes, that the same metric also gives an exact description
of a patch of AdS space. In any case we will call these the
half--plane coordinates. The two dimensionless parameters $R/l_s$
and $g_s$ of the bulk theory -- $l_s$ is the string length scale --
are related to the SYM quantities $N$ and $g$ by
\begin{eqnarray}
R/l_s&=&(g^2 N)^{1/4} \cr g_s &=& g^2.
\end{eqnarray}
The 5 and 10 dimensional Newton constants are given by
\begin{eqnarray}
G_5 &=& G_{10}/R^5 \cr G_{10}&=& g_s^2 l_s^8.
\end{eqnarray}
We set $R = 1$ for simplicity.
The string length scale is then given by $l_s = 1/(g^2 N)^{1/4}$.
Throughout we neglect numerical factors of order unity.


\setcounter{equation}{0}
\section{Bulk Waves}

As in \cite{4}, we model bulk waves with ``squeezed states'' in the
boundary theory.
Our goal is to study expectation values of various
operators in the ``squeezed states'' and
identify the precursors that store local bulk information.

For definiteness, let us consider a gravitational wave propagating
radially outward from $r \sim 0$.
In the next section, we will be interested in the case of a dilaton wave.
Assume that the wave is in one of the lowest spherical harmonics on
the 3--sphere.
In half--plane coordinates the plane fronted wave has the form
\begin{equation}
\gamma_{\mu \nu}(z,x,t) = \xi_{\mu\nu} \sqrt{G_5} {\Phi(z,t)\over z^2},
\end{equation}
where $\gamma_{\mu \nu}(z,x,t)$ is defined by
\begin{equation}
ds^2 = \left[{1\over z^2}(dt^2 -dz^2 - dx^i dx^i) \right]  +  \gamma_{\mu
\nu}(z,x,t)dx^{\mu}dx^{\nu}
\end{equation}
and $\xi_{\mu\nu}$ is a transverse traceless polarization tensor
with non--vanishing components in the $x$ directions. The
polarization tensor is assumed normalized to unity.

Far away from the original sources, $\Phi(z,t)$ satisfies the same wave
equation as a free, minimally coupled, massless scalar field in AdS.
We use normalization conventions so that $\Phi(z,t)$ is canonically
normalized.
Thus we keep the amplitude $|\Phi(z,t)|$ finite, independent of N, and
the energy of the wave is finite.
The corresponding operator in the SYM theory is $\xi_{ij}T_{ij}/N$.
The 2-point function of this operator is of order $N^0$.
Non-linear terms in the gravitational field equations are
suppressed by additional factors of $\sqrt{G_5} \sim N^{-1}$ and will
be ignored in this
paper \footnote { In \cite {4}, these effects were included and it was
shown how they can be reproduced in free field theory up to possible
numerical coefficients.
In the large N limit they are suppressed. However, they are important
to recover consistency in the behavior of the energy momentum tensor
required by general principles of quantum mechanics.
We refer the reader in refs. \cite{4}\cite{10} for a discussion of this
point.}.

Near the boundary, normalizable solutions to the wave equation behave as
follows
\begin{equation}
\Phi (z, t) \sim z^4 \int{d\omega |\omega|^3  \phi(\omega) e^{-i\omega
t}},
\end{equation}
with $\phi(\omega) = \phi^*(-\omega)$ since the field is real.
According to the AdS--CFT correspondence, the wave makes a
contribution to the SYM energy momentum tensor given by \cite{11}
\begin{equation}
\langle{T_{ij}\over N}\rangle \sim - \xi_{ij} z^{-4} \Phi(z,t)|_{z=0}.
\end{equation}

We are interested in describing a wave emitted at a particular time $t_0$ in
the past, near $r \sim 0$, so that, when $t < 0$, the perturbation
vanishes within a neighborhood of the boundary.
This can be achieved if we choose the function $|\omega|^3
\phi(\omega)$ to be analytic in
the upper half $\omega$--plane and have the right asymptotic behavior as
$\omega
\rightarrow i \infty$. Then the boundary data vanish for $t < 0$ and so
does the
contribution to $<{T/ N}>$.
In general, the boundary data will be non-vanishing when $t > 0$ since
$\phi(\omega)$ will have singularities in the lower half-plane.
Also, causality of the bulk theory insures that the function
$\Phi(z,t)$ describes a wave which, at any $ t_0 < t < 0$, exactly vanishes
for $z < |t|$.
In addition, bulk causality requires that all local bulk fields
evaluated in a neighborhood of the boundary, as
well as products of such fields, retain their vacuum expectation
values until $t = 0$.
Therefore, on the SYM side, expectation values of local gauge
invariant operators and their products must be identical to their
vacuum values until $t = 0$.

\bigskip
\noindent{ \bf Squeezed States in Yang--Mills Theory }

We propose that during the propagation of the wave, the SYM theory is
excited in the ``squeezed state'' defined by
\begin{equation}
\left| \Psi \right> = \exp { \left[{ i \xi_{ij} \over N} \int{d^3 \vec{x}
dt f(t) T_{ij} (\vec{x},t)} \right] } \left| \Omega \right>,
\end{equation}
where $|\Omega\rangle$ is the vacuum of the interacting theory and $f(t)$ is
some real function related to the boundary data of the
wave.
The polarization $\xi_{ij}$ is taken to be traceless and symmetric.
It will turn out to be the polarization of the wave.
The state thus defined is unit--normalized \footnote{In the free theory, to
leading order in $1/N$, it reduces to the
``squeezed state'' considered in \cite{4}, up to some normalization
factor. Also, the energy momentum tensor is normal--ordered so
that the vacuum energy density is zero.}.

Our motivation in writing Eq(2.5) is as follows.
In the large N limit with the 't Hooft coupling held fixed and large,
$|\Psi\rangle$ corresponds to a coherent state in the bulk.
To see this note that if we Fourier expand any local gauge invariant operator
$O(\vec{x},t)$
\begin{equation}
O(\vec{x},t) = \int_{\omega > 0}{d\omega d^3 \vec{k}
{\cal{O}}(\omega,\vec{k}) e^{-i
\omega t + i \vec{x}\cdot \vec{k}} +
h.c.},
\end{equation}
then, to leading order in $1/N$, the positive frequency modes
${\cal{O}}(\omega,\vec{k})$ behave like annihilation operators and the
negative frequency modes
${\cal{O}}^{\dagger}(\omega,\vec{k})$ behave like creation
operators \cite{13}.
In particular, their commutator is a c--number function.
Thus, up to some irrelevant normalization factor, the state $|\Psi
\rangle$ takes the form
\begin{equation}
\left| \Psi \right> \sim \exp { \left[  \int{
d\omega f(\omega) {\cal{O}}^{\dagger}(\omega,0) } \right] } \left|
\Omega \right>.
\end{equation}
We see that if we identify the SYM
vacuum with the bulk vacuum and the modes ${\cal{O}}(\omega,\vec{k})$
with the Fourier modes of the bulk field corresponding to $O$,
$\left| \Psi \right>$ becomes a coherent bulk state.
This can always be done in the limit we are considering since in this
limit, the relation
\begin{equation}
O(x) = z^{-4}\Phi(z,x)|_{z \rightarrow 0}
\end{equation}
holds as an exact operator relation \cite{12}\cite{13}.
Coherent states in the bulk describe classical waves.

Next we calculate the expectation value
\begin{equation}
\left<\Psi\right| {T_{ij}(\vec{y},\tau)\over N}  \left| \Psi \right>
= {1 \over N}\left < \Omega \right| e^ {-{ i \xi_{mn} \over N} \int{ f(t)
T_{mn}(x)} }
{T_{ij}(y)} e^ { { i \xi_{mn} \over N} \int{ f(t) T_{mn}(x)}
} \left| \Omega \right >.
\end{equation}
In the 't Hooft limit and to leading order in $1/N$, the commutator
$\left[ T_{ij}(y),T_{mn}(x) \right]$ is a c--number function
proportional to the central charge of the theory.
Therefore, it is of order $N^2$.
In fact, it is independent of the 't Hooft coupling and can be
calculated in the free theory.
This function vanishes both inside and outside the light-cone; it
receives contributions only when the points $x$ and $y$ are at
light--like separation.
Hence, we can commute $T(y)$ past the exponential picking a factor
proportional to this commutator.
Recall also that the energy momentum tensor has zero expectation value
in the vacuum.
Then, to leading order in $1/N$, the expectation value is given by
\begin{equation}
\left<\Psi\right| {T_{ij}(\vec{y},\tau) \over N}  \left| \Psi \right>
= {i\xi_{mn} \over N^2}\int {dtd^3\vec{x}f(t) \left[
T_{ij}(\vec{y},\tau),T_{mn}(\vec{x},t) \right]}.
\end{equation}
The commutator is determined by the imaginary part of the time-ordered
2-point function of the energy momentum tensor, and so
\begin{equation}
\left<\Psi\right| {T_{ij}(\vec{y},\tau) \over N}  \left| \Psi \right>
= -\xi_{mn}{2 \over N^2}\int {dtf(t)\epsilon(\tau-t)
Im\int d^3\vec{x}\left< T_{ij}(\vec{y},\tau)T_{mn}(\vec{x},t) \right>} +
O({1 \over N}).
\end{equation}
The expectation value inside the integral is in the vacuum.
All other components of the energy momentum tensor have 
expectation values of order $1/N$ in this state.

The details of the calculation can be found in the appendix.
Here, we write down the results.
The spatial integral is imaginary and independent of $\vec{y}$.
Simple dimensional analysis shows that it behaves like
\begin{equation}
{1 \over |\tau - t|^5}.
\end{equation}
Thus the expectation value of the energy momentum tensor in the
``squeezed state'' is given by
\begin{equation}
\left<\Psi\right| {T_{ij}(\vec{y},\tau) \over N}  \left| \Psi \right>
\sim \xi_{ij} \int {dtf(t) {1 \over (\tau - t)^5}}.
\end{equation}
If we Fourier transform $f(t)$
\begin{equation}
f(t) = \int{d\omega {f(\omega) \over \omega} e^{-i \omega t}},
\end{equation}
with $f(\omega) = -f^{*}(- \omega)$, we obtain
\begin{equation}
\left<\Psi\right| {T_{ij}(\vec{y},\tau) \over N}  \left| \Psi \right>
\sim i\xi_{ij} \int {d\omega  f(\omega) |\omega|^3 e^{-i \omega \tau}}.
\end{equation}
Comparing with Eq(2.3), we must set
\begin{equation}
\phi(\omega) \sim i f(\omega).
\end{equation}
Note that we have chosen $\phi(\omega)$ so that the boundary data
vanish for $t < 0$.
This does not imply that $f(t)$ is zero for $t < 0$.

Finally, consider expectation values of products of the energy
momentum tensor.
Using the same method as before, we can easily see that these will
differ from their vacuum values only by products of commutators.
Schematically, we have
\begin{equation}
{1\over N^2}\left<\Psi\right| {T_1 }{T_2 }  \left| \Psi \right>
= {1 \over N^2}\left< \Omega \right|{T_1 }{T_2 }\left|\Omega\right> -
{1 \over N^2}\int{ \left[T_2,T\right]} \int{\left[T_1,T\right]}.
\end{equation}
The non-trivial component is just the product $\left<\Psi\right| {T_1
}  \left| \Psi \right>\left<\Psi\right| {T_2 }  \left| \Psi
\right>$. Therefore, for $t_1, t_2 < 0$, the expectation value is identical to
its vacuum value since each factor vanishes by construction.
In any case, products of local gauge invariant operators contain no additional
information about the profile of the wave.
This is of course a consequence of large $N$ factorization.


\setcounter{equation}{0}
\section{Wilson Loops}

In this section we show how the expectation value of a Wilson loop
in a squeezed state carries non-trivial information about a
dilaton wave. Since we are interested in the instantaneous state
vector, a Wilson loop will typically mean a spatial loop with no
extension in the time direction.
To model a dilaton wave in the SYM theory, we replace
$\xi_{ij}T_{ij}/N$ with $O = TrF^2/N$ in Eq(2.5).

We consider a conventional Wilson loop
\begin{equation}
W({\cal{C}}) = Tr P e^{i \oint {A_{\mu}dx^{\mu}}}
\end{equation}
for simplicity.
In the 't Hooft limit and large 't Hooft coupling, the
vacuum expectation value of this loop can be obtained from the area of
a minimal world-sheet in AdS that ends on the loop at the boundary
\cite{8}.
We consider a spatial Wilson loop evaluated at $ \tau
< 0$ and oriented in the $x_1 - x_2$ plane.
We take the loop to be circular with size $a$. We regularize the VEV
of this loop by dividing the divergent term proportional to the
circumference.

We wish to calculate the expectation value $\left<\Psi\right| {W}
\left| \Psi \right>$ in the case when $f$ is small.
In this case, we can expand the exponential keeping only linear terms
in $f$.
We do not expect higher order terms to modify our conclusions
significantly, since in the 't Hooft limit their expectation values
should factorize into products involving the linear term together with
featureless (independent of $\tau$ and $a$) factors such as the VEV of
products of $O$.

Then the expectation value reduces to the following expression
involving the commutator of the loop with the operator $O$
\begin{equation}
\left<\Psi\right| {W}\left| \Psi \right> =
\left<W\right> + i \int {dtd^3\vec{x}f(t)\left< \left[ W(\tau),O(\vec{x},t)
\right]\right>}.
\end{equation}
All expectation values in the RHS of the equation are vacuum
expectation values.
The first term is irrelevant to us since, by conformal invariance, it
should be independent of the size of the loop $a$ (and $\tau$).
In terms of time--ordered vacuum expectation values the second term
takes the form
\begin{equation}
i \int {dtd^3\vec{x}f(t) \epsilon(\tau-t) \left[\left<
W(\tau)O(\vec{x},t)\right> - \left<
W^{\dagger}(\tau)O(\vec{x},t)\right>^{*}\right] }.
\end{equation}
The hermitian conjugate of the loop operator is obtained by reversing
the orientation of the loop in the $x_1-x_2$ plane.

The Euclidean version of the ``2-point functions'' appearing in Eq(3.3) has
been computed in \cite{9} using the correspondence.
One first finds a minimal world-sheet with the loop as its boundary.
The world-sheet in turn induces a source term in the dilaton field
equations through the coupling
\begin{equation}
{1 \over 2 \pi \alpha'} \int{d^2\sigma \sqrt{h}} e^{\phi \over 2}.
\end{equation}
Here, $h_{ij}$ is the metric induced on the world-sheet when the
background metric is in the Einstein frame.
The term in the world-sheet action involving the curvature of the
world-sheet is
suppressed when the 't Hooft coupling is large and can be ignored.
The 2-point function is given by the boundary data of the
dilaton profile obtained by solving the classical field equations in
the presence of the source.
It depends only on two parameters, which are the polar co-ordinate of
the operator $O$ on the plane defined by the loop $r$ and its
perpendicular distance from the plane of the loop $ y =
\sqrt{(t-\tau)^2 + x_3^2} $ \cite{9}:
\begin{equation}
\left< W(\tau)O(\vec{x},t)\right> \sim {\left< W\right> \over N}
{a^4 \over {\left[ {(y^2 + r^2 - a^2)}^2 + 4a^2y^2 \right]}^2}.
\end{equation}
We see that the 2-point function behaves like
$1/d^4$ when the operator approaches the loop, where $d = \sqrt{ y^2 +
(r-a)^2}$ is the distance of the operator from the loop.
To obtain the expression in Minkowski signature, we replace
$(\tau - t)^2 \rightarrow -(\tau - t)^2 + i\epsilon$.

Before we continue with our calculation, we make some remarks about
this correlation function.
First, we see that it is of order $N^0$ since the expectation value of
the loop itself is of order $N$.
In fact, we may think of the operator $O = TrF^2 / N$ as a small
Wilson loop.
The disconnected part of the 2-point function is zero since $O$ has
vanishing VEV.
The connected part of the 2-point function receives contributions from
world-sheets in the bulk that have the two loops as boundaries.
The topology of these surfaces implies that the 2-point
function is of order zero in the large $N$ expansion.
Second, reversing the orientation of the loop does not change the
result for the dilaton profile since the coupling of the
world-sheet in the bulk to the dilaton field, Eq(3.4), remains the same.
Hence, Eq(3.2) reduces to the following expression
\begin{equation}
\left<\Psi\right| {W}\left| \Psi \right> =
\left<W\right> - 2 \int {dtf(t)\epsilon(\tau - t)Im \int d^3\vec{x} \left<
W(\tau)O(\vec{x},t)\right>}.
\end{equation}

First we do the spatial integral over the 2-point function and obtain
the imaginary part as a function of the ratio
\begin{equation}
\lambda = {|\tau - t| \over a}.
\end{equation}
We also rescale $x_3$ and $r$ so that the variables of integration are
dimensionless.
In polar co-ordinates the integral takes the form
\begin{equation}
I = {2\pi \over a}\int{drdx_3{r \over \left[ x_3^2 - A^2 +
i\epsilon \right]^2 \left[x_3^2 - B^2 + i\epsilon\right]^2}},
\end{equation}
where
\begin{equation}
A^2  = {\lambda^2 - (r - 1)^2}
\end{equation}
and
\begin{equation}
B^2 = {\lambda^2 - (r + 1)^2}.
\end{equation}
The integrand has poles when $A^2$ and $B^2$ are positive.
Therefore the integral has non-vanishing imaginary part.
We explain the physical origin of these poles at the end of this section.
We integrate over $x_3$ first, closing the contour from below and
picking up the residues at the poles in the lower-half plane.
Only non-negative real poles contribute to the imaginary part as a result
of the $i\epsilon$ prescription.

In the appendix, we analyze the behavior of the imaginary part of the
integral for three cases.
When $\lambda \gg 1$, we find
\begin{equation}
Im(I) \sim {1 \over a\lambda^5} = {a^4 \over |\tau - t|^5}.
\end{equation}
The result is identical to the result found in Eq.(2.12) for the case of
local operators.
This is of course the behavior one should expect to see.
In this case, the temporal separation between the loop and the operator
O is much bigger than the size of the loop, and we should be able to
use the operator product expansion of the loop in terms of local gauge
invariant operators to calculate the 2-point function.
Note also that the 2-point function behaves like
\begin{equation}
\left<WO\right> \sim {a^4 \over \left[x^2 - (\tau - t)^2\right]^4}
\end{equation}
when $\lambda \gg 1$, as the 2-point function of $O$ with itself.
As $\lambda \rightarrow 1$, the imaginary part increases.
When $\lambda \sim 1$, it is the biggest
and behaves like
\begin{equation}
Im(I) \sim {1 \over a( \lambda - 1 )^{3/ 2}}.
\end{equation}

When $\lambda \ll 1$, we find that the imaginary part tends
to zero like
\begin{equation}
Im(I) \sim  {\lambda^2 \over a}.
\end{equation}
We can understand the result as follows.
As explained below, in this case, the imaginary part of the integral
receives contributions only when the operator is very close to the
loop at $r \sim 1$ and $x_3 \sim 0$.
Their temporal separation is also small.
Thus, using the Heisenberg equations of motion, we can approximate
$O(\vec{x},t)$ with
\begin{equation}
O(\vec{x},t) = O(\vec{x},\tau) - \partial_tO(\vec{x},t)|_{t=\tau}(\tau
-t).
\end{equation}
We see that the operator $O$ commutes with the Wilson loop unless the two are
in contact.
Essentially, only a single point of the loop contributes to the
commutator, a measure zero effect.
The commutator in turn determines the imaginary part of the integral
as we can see from Eq(3.2) and Eq(3.3). So we expect the imaginary part of the
integral to vanish like a power of $\lambda^2$ or faster.

Let us now see how the expectation value of the Wilson loop in the
``squeezed state''
\begin{equation}
\left<\Psi|W(\tau)|\Psi\right> = -2 \int dt f(t)\epsilon(\tau - t) Im(I)
\end{equation}
carries information about the corresponding dilaton wave.
The imaginary part of the integral is a function of $\lambda = |\tau -t|/a$.
As before, we choose $f(t)$ so that $ \left<\Psi|O(\tau)|\Psi\right>$ exactly
vanishes when $\tau < 0$. At any $\tau < 0$, the
corresponding bulk wave vanishes for $z < |\tau|$.
On the other hand, the expectation value of the Wilson loop has non-trivial
time dependence when $\tau < 0$.
Early in the remote past, when $|\tau| \gg a$, we can approximate
$Im(I) \sim 1/|\tau - t|^5$.
Therefore, the expectation value tends to zero since it behaves
exactly the same way as the expectation value of local gauge
invariant operators given in Eq(2.13).
When $|\tau| \ll a$, the imaginary part of $I$ is essentially
independent of $\tau$ within most of the domain of integration but a
small interval when $|t| \sim |\tau|$.
Thus the expectation value receives its time--dependence from this small
region of integration.
Within this region though, $\lambda \ll 1$ and so the imaginary
part of $I$ is tiny.
Hence, the expectation value is featureless having essentially no
time--dependence.
When $|\tau| \sim a$, the expectation value receives non-trivial
time--dependence due to competition effects between $f(t)$ and the imaginary
part of $I$.
It receives its biggest contribution from the region of integration
near $t \sim 0$ since then $\lambda \sim 1$ and the imaginary part of
$I$ diverges.
When $|\tau| \sim a$, the wave is at co-ordinate distance $\sim a$ from the
boundary.
Thus the Wilson loop ``detects'' the wave when its distance from the
boundary is comparable to the size of the loop, and reproduces
details that depend on the profile of the wave.
This is of course a manifestation of the UV--IR relation \cite{3}.

Another interesting example, is the case when $f(t)$ is oscillatory
near $t = 0$ and exponentially small otherwise. The oscillations are
well concentrated near $t = 0$. For example, we may take $f(t)$ to be a
polynomial in $t$ times a gaussian. At any time $\tau$ other than
zero, the corresponding bulk wave should be oscillatory near $z = |\tau|$
and very small in a neighborhood of the boundary. In this case,
expectation values of local gauge invariant operators behave like
\begin{equation}
\left<\Psi|O(\tau)|\Psi\right> \sim {f(0)\delta t \over |\tau|^5}
\end{equation}
and so they remain small unless the wave is at the boundary at $\tau
= 0$. Here, $\delta t$ is the characteristic decay time of the
oscillations in $f(t)$. The expectation value of the Wilson loop
though has very different time--dependence. Again, using the results
for the behavior of the imaginary part of $I$ as a function of the
ratio $|\tau|/a$, one can see that the expectation value is
oscillatory when $|\tau| \sim a$ and suppressed when $|\tau| \gg a$ or
$|\tau| \ll a$.

In short, when the wave is very close to the boundary, only small Wilson loops
are excited. At that time, however, expectation values of local gauge
invariant
operators begin to oscillate. On the other hand, when the wave is far
from the boundary only big Wilson are excited. This shows that the
precursors are in fact Wilson loops.

\

Finally, let us try and understand the physical origin of the poles in
the integrand in Eq(3.8).
When the denominators vanish, the 2-point function has non-vanishing
imaginary part since then the $i\epsilon$ prescription for
treating the poles becomes relevant.
As one can see from Eq(3.2) and Eq(3.3), the imaginary part of the 2-point
function is determined by the vacuum expectation value of the commutator
between the Wilson loop and the operator $O$.
Therefore, at the poles the commutator is non-vanishing.
Now, the commutator can be non-zero only when some part of the loop of
non-trivial measure is on the light-cone of $O$. Then the
commutator between the vector potential at any point whose separation
from $O$ is light-like and $O$ is non-zero, and in turn all of them contribute
to the commutator between the Wilson loop itself and $O$.
This is precisely what happens at the poles as we show below.
The imaginary part of the 2-point function vanishes when the loop is
not intersecting the light-cone, and the contribution to the integral
from this region of integration is real.
Then, the 2-point function is non-singular as well.

Suppose the operator is at $t=0$.
Then the loop can intersect with the past light-cone of $O$ only.
For $|x_3| < \lambda$, the light-cone intersects the $x_1 - x_2$ plane
at a circle of radius
\begin{equation}
\rho = \sqrt{\lambda^2 - x_3^2}.
\end{equation}
The polar co-ordinate $r$ of $O$ is the distance of the center of this
circle from the center of the loop.
The point of the loop closest to the center of the light-cone circle is
at distance $|r-1|$ from it, while the one that is the farthest is at
distance $r+1$.
Clearly, when $A^2$ is negative, the loop is outside the light-cone and
so no contributions to the imaginary part of the integral arise from
this region of integration for any $\lambda$.

When $A^2$ is positive the loop and the light-cone circle intersect.
We may choose, however, $|x_3| = A$ so that the two circles are
tangent to each other.
This is precisely when the integrand is singular.
When the two circles are tangent the set of points on the
loop that are close enough to the light-cone is of bigger measure and we get a
contribution to the commutator and a pole in the 2-point function.
For $\lambda > 1$ the light-cone circle is tangent to the loop from the
outside. The opposite is true for $\lambda < 1$. In this case, the light-cone
circle becomes smaller and smaller as $\lambda \rightarrow 0$ and the
effect ceases to be important.

For $\lambda \geq 1$, we can choose $x_3$ small enough so that $\rho$ is
bigger than the radius of the loop.
If $B^2$ is positive, then, for $|x_3| < B$, the loop is inside the
light-cone.
For $|x_3| = B$ the two circles are tangent and again we have a pole
in the 2-point function.

For $\lambda \geq 1$, the 2-point function becomes even more singular when
$A =
B$ at $r = 0$ and $\rho = 1$.
In this case, the whole loop is on the light-cone.
Therefore, we should expect a big contribution to the imaginary
part of the integral from the small $r$ region.
We expect this effect to amplify when $\lambda \sim 1$ since then, $x_3
\sim 0$, and the operator is closer to the plane of the loop.
Note also that the 2-point function is as singular when $A$ or $B$ are
zero and $x_3 = 0$.
The two effects combine when $\lambda = 1$.
The 2-point function can be the most singular in this case and we expect
the imaginary part of the integral to be the biggest.

\setcounter{equation}{0}
\section{Discussion}

The main purpose of this paper is to identify the non-local precursor
fields of the SYM boundary theory that record information about
local processes occurring deep in the interior of the bulk AdS
spacetime.
Causality of the bulk theory requires that the precursors are
intrinsically non-local. They are not simple
products of local operators corresponding to the classical supergravity
fields in the standard AdS--CFT dictionary. Correlation functions of
such products essentially remain featureless until the
signal from the event arrives at the boundary. Yet, as in \cite{4}, we
argue that the precursors store the
information long before the signal can propagate to the boundary.

In this paper, we study a rather simple case involving the propagation of a
classical bulk wave toward the boundary. It is shown that when the
wave vanishes within a neighborhood of the boundary, products of local
gauge invariant operators retain their vacuum expectation values,
whereas Wilson loops are excited when their size is of the same order as
the co-ordinate distance of the wave from the boundary.
A detailed translation of all the configurations of the bulk theory to the SYM
theory is not yet available, but, as in the example of the wave, we
believe that the precursors will involve Wilson loops with size
dictated by the UV--IR connection.
The precise way Wilson loops would store information about
complicated processes in the bulk is very difficult to see. In particular,
it remains a challenge to understand what precursors describe small
Schwarzschild black holes at the center of AdS, or what configurations
of Wilson loops provide the signal that a black hole forms in a
head-on collision of two very energetic gravitons \cite{4}.
However, Wilson loops and their products form a
complete set of gauge invariant operators in the SYM theory. This
means that at any time one should be able to recover all the
information about the state of the theory from their expectation
values and expectation values of their products.

Our analysis has been carried out in the 't Hooft limit where we keep
the 't Hooft coupling fixed and large and take $N \rightarrow \infty$.
In this limit the bulk theory is manifestly local as it is well
described by linearized supergravity. We do not consider $1/N$
corrections in this paper since they are too small.
We think that their effect is to modify the original
expectation values of local gauge invariant operators by
featureless components that do not carry any interesting information
about the details of the relevant bulk process. For example, in the
case of the gravitational wave considered in section II, the
next--to--leading order $1/N$
corrections depend on the total energy in the bulk, which is
constant, but not on the detailed profile of the wave.
Any interesting effect of bulk interactions should be recovered from
such expectation values only when the signal of the event arrives at
the boundary.

We believe that the ``squeezed states'' constructed in the SYM theory
continue to accurately describe gravitational waves including the
effect of bulk interactions.
Evidence for this was found in \cite{4}, where the success of the free
field theory model considered was linked with the non-renormalization
theorem for the 3-point function of the energy momentum tensor.
It would be very interesting to study the exact description of a
gravitational wave in the flat space limit considered in
\cite{15}\cite{16}.
In this limit we take $N$ large and $g$ fixed. We also keep bulk
energies fixed in string units. This means that we have to consider
energies in the SYM theory that scale like $N^{1/4}$.
In flat space, plane gravitational waves are exact solutions of the
theory and do not receive any stringy corrections \cite{17}.
However, we do not have any computational control in the SYM theory
in this limit apart from conjectured non-renormalization theorems for
the 2-point and 3-point functions of chiral primaries.

What really distinguishes the precursors in the case of ${\cal{N}}=4$
SYM theory from other non-local observables in the theory is that
Wilson loops cannot be expressed in terms of finite polynomials of
local gauge invariant operators corresponding to the bulk fields.
Gauge invariance equips the boundary holographic theory with this rich
class of
intrinsically non-local observables so that it can reproduce traces of bulk
causality and locality. Thus gauge invariance is crucial in the way
this particular local conformal theory describes
bulk physics. It would be interesting to understand the precise nature
of the precursors in other AdS--CFT dualities in which the CFT is not
a conventional gauge theory. For example, the $AdS_3$ case.\footnote
{Some interesting issues
concerning this particular case were recently studied in \cite{14}.}
In some of these examples the CFT is
obtained from a gauge theory through renormalization group flows; however
there is no remnant of the original gauge symmetry at the fixed point.
It is particularly challenging to find special non-local observables
in these examples as well so as to understand better the holographic
nature of gravity.

\

{\bf Acknowledgements }
We would like to thank Vijay Balasubramanian, Gary Felder, Jason
Prince Kumar, Maxim Perelstein, Joseph Polchinski, Simon Ross, Steve
Shenker and Eva Silverstein for useful discussions.
This work was supported in part by NSF grant 9870115 and by the US
Department of Energy under contract DE--AC03--76SF00515.




\appendix

\section*{Appendix A}
\setcounter{section}{1}

In this appendix, we show how to compute the expectation value of the
energy momentum tensor in the ``squeezed state'' as given in Eq(2.11):
\begin{equation}
\left<\Psi\right| {T_{ij}(\vec{y},\tau) \over N}  \left| \Psi \right>
= -\xi_{mn}{2 \over N^2}\int {dtf(t)\epsilon(\tau-t)
Im\int d^3\vec{x}\left< T_{ij}(\vec{y},\tau)T_{mn}(\vec{x},t) \right>}.
\end{equation}
The 2-point function can be found in \cite{6}.
It is given by
\begin{equation}
\left< T_{ij}(\vec{y},\tau)T_{mn}(\vec{x},t) \right> \sim
N^2 X_{ijmn} {1 \over \left[(\vec{x} - \vec{y})^2 - (\tau - t)^2 +
i\epsilon \right]^2},
\end{equation}
where we drop numerical factors of order unity.
The tensor $X_{ijmn}$ involves terms with four derivatives with
respect to $y$.
The precise formula can be found in \cite{6}.

Next we calculate the integral
\begin{equation}
\int d^3\vec{x} {1 \over \left[(\vec{x} - \vec{y})^2 - (\tau - t)^2 +
i\epsilon \right]^2}
\end{equation}
and obtain its imaginary part. The integral is independent of
$\vec{y}$. We can also scale $|\Delta t| = |\tau - t|$ out of the
integral to obtain the following expression
\begin{equation}
{4\pi I \over |\Delta t|},
\end{equation}
where
\begin{equation}
I = \int_0^\infty dx {x^2 \over \left(x^2 - 1 + i\epsilon\right)^2}.
\end{equation}
Using integration by parts we can simplify the integral as follows
\begin{equation}
I = {1 \over 4}\int_{-\infty}^\infty dx {1 \over \left(x^2 - 1 +
i\epsilon\right)}.
\end{equation}
This integral can be done by contour integration. We close the contour
from below picking up the residue at the pole $x = 1 - i\epsilon/2$.
We find
\begin{equation}
I = - {i \pi \over 4}.
\end{equation}
Since the integral is independent of $\vec{y}$ only the term with four
time--derivatives in $X_{ijmn}$ contributes to the expectation value.
Then the expectation value reduces to the following expression
\begin{equation}
\xi_{ij} \int dt f(t) \epsilon(\tau - t) \partial_{\tau}^4 \left({1
\over |\tau - t|} \right).
\end{equation}
This is the same expression as equation Eq(2.13) in Section II.





\setcounter{section}{2}
\setcounter{equation}{0}
\section*{Appendix B}

In this appendix, we show how to compute the imaginary part of the integral
\begin{equation}
I = {2\pi \over a}\int{dx_3dr{r \over \left[ x_3^2 - A^2 +
i\epsilon \right]^2 \left[x_3^2 - B^2 + i\epsilon\right]^2}},
\end{equation}
where $A^2 = \lambda^2 - (r - 1)^2$ and $B^2 = \lambda^2 - (r + 1)^2$
as defined in Section III.
We first do the $x_3$--integration closing the contour from below and
evaluating the residue at the poles.
Only real poles contribute to the imaginary part of the integral.
We study the cases $\lambda > 1$ and $\lambda < 1$ separately.
For $\lambda > 1$, $A^2$ is positive for $ 0 < r < \lambda +1$ and
$B^2$ is positive for $ 0 < r < \lambda - 1$.
For $\lambda < 1$, $A^2$ is positive for $ -\lambda +1 < r < \lambda
+1$ while $B^2$ is negative for all values of $r$.

For $\lambda > 1$, the imaginary part of I is obtained from the
imaginary part of the following expression
\begin{eqnarray}
&&
{i\pi^2 \over 16a}\int_0^{\lambda - 1}{dr{1 \over r} \left[ {1 \over (B
- i\delta)^3} - {1 \over r}{1 \over (B - i\delta)}\right]} \nonumber\\
&& + \ \
{i\pi^2
\over 16a}
\int_0^{\lambda + 1}{dr{1 \over r}\left[ {1 \over (A
- i\delta)^3} +  {1 \over r}{1 \over (A - i\delta)}\right]}.
\end{eqnarray}
This expression is obtained after we calculate the residue at the poles
\begin{equation}
x_3 = A - i\delta
\end{equation}
and
\begin{equation}
x_3 = B - i\delta.
\end{equation}
Here, $\delta$ is a small number to be set to zero at the end of the
calculation.

Let us obtain the imaginary part for the case when $\lambda \gg 1$
first.
We show that it vanishes like $1/\lambda^5$.
We split the integrals into three pieces:
\begin{equation}
I_1 = {i\pi^2 \over 16a}\int_0^{1}{dr{1 \over r} \left[ {1 \over B^3} +
{1 \over A^3} - {1 \over r} \left( {1 \over B} - {1 \over A}\right)\right]},
\end{equation}
\begin{equation}
I_2 ={ i\pi^2 \over 16a}\int_1^{\lambda - 1}{dr{1 \over r} \left[ {1 \over (B
- i\delta)^3} - {1 \over r}{1 \over (B - i\delta)}\right]},
\end{equation}
and
\begin{equation}
I_3 ={ i\pi^2 \over 16a}\int_1^{\lambda + 1}{dr{1 \over r} \left[ {1 \over (A
- i\delta)^3} + {1 \over r}{1 \over (A - i\delta)}\right]}.
\end{equation}

The integrand in $I_1$ looks singular at $r = 0$, but in fact it
behaves like $r^0$ as $r \rightarrow 0$. To see
this, we Taylor expand $A$ and $B$ in powers of $r$ to obtain
\begin{equation}
{1 \over B} - {1 \over A} = {2r \over (\lambda^2 - 1)^{3/2}} +
O\left({r^2 \over (\lambda^2 - 1)^{5 / 2}}\right),
\end{equation}
and
\begin{equation}
{1 \over B^3} + {1 \over A^3} = {2 \over (\lambda^2 - 1)^{3 / 2}} +
O\left({r \over (\lambda^2 - 1)^{5/ 2}}\right).
\end{equation}
This can be done since $0 < r <1$ and $\lambda \gg1$.
Thus the integrand is of order $1/\lambda^5$, and, therefore,
\begin{equation}
Im(I_1) \sim {1 \over a\lambda^5}.
\end{equation}

Next, consider $I_2$. The integrand is singular at $r = \lambda -1$
when $B = 0$, but as we will show the imaginary part is finite. The
small number $\delta$ regulates the imaginary part.
The imaginary part is given by
\begin{equation}
Im(I_2) ={\pi^2 \over 16a}\int_1^{\lambda - 1}{dr{1 \over r} \left[
{(B^3 - 3\delta^2B) \over (B^2
+\delta^2)^3} - {1 \over r}{B \over (B^2 + \delta^2)}\right]}.
\end{equation}
Change variables by setting
\begin{equation}
r + 1 = \sqrt{\lambda^2 - x^2}
\end{equation}
to find
\begin{eqnarray}
Im(I_2) &=& {\pi^2 \over 16a}\int_0^{\sqrt{\lambda^2 - 4}}dx \left\{ {1
\over (\lambda^2 - x^2)^{1 / 2}\left[(\lambda^2 - x^2)^{1 /
2}-1\right] } \right. \times \nonumber \\
& &\left. \left[
{(x^4 - 3\delta^2x^2) \over (x^2
+\delta^2)^3} - {1 \over\left[(\lambda^2 - x^2)^{1/ 2}-1\right]
}{x^2 \over (x^2 + \delta^2)}\right] \right\}.
\end{eqnarray}
Similarly, if we change variables
\begin{equation}
r - 1 = \sqrt{\lambda^2 - x^2},
\end{equation}
the imaginary part of $I_3$ becomes
\begin{eqnarray}
Im(I_3) & = & {\pi^2 \over 16a}\int_0^{\lambda} dx \left\{ {1
\over (\lambda^2 - x^2)^{1/ 2} \left[ (\lambda^2 - x^2)^{1 /
2}+1 \right]}\right. \times \nonumber \\
&  & \left. \left[
{(x^4 - 3\delta^2x^2) \over (x^2 + \delta^2)^3}
+ {1 \over\left[(\lambda^2 - x^2)^{1 / 2}+1\right]  }
{x^2 \over (x^2 + \delta^2)}\right]\right\}.
\end{eqnarray}

Combining the two we are left with the following simpler integrals
\begin{equation}
X_1 = \int_0^{\sqrt{\lambda^2 - 4}}{dx{2
\over \left(\lambda^2 - 1 -  x^2 \right) } \left[
{(x^4 - 3\delta^2x^2) \over (x^2
+\delta^2)^3} - {2 \over\left(\lambda^2 - 1 -  x^2\right)  }{x^2 \over (x^2
+ \delta^2)}\right]},
\end{equation}
and
\begin{equation}
X_2 = \int_{\sqrt{\lambda^2 - 4}}^\lambda{dx{1
\over (\lambda^2 - x^2)^{1/ 2}\left[(\lambda^2 - x^2)^{1 / 2}+1\right] }
\left[
{1 \over x^2} + {1 \over\left[(\lambda^2 - x^2)^{1 / 2}+1\right]
}\right]}.\end{equation}
In $X_2$ we drop the terms proportional to $\delta$
since the integrand is well behaved within the domain of integration. This
integral can be obtained in terms of logarithms. We do not write the
whole expression down. Rather, we write its series expansion in terms
of powers of $1/\lambda$:
\begin{equation}
X_2 = {2 \over 3\lambda} + {4 \over 3\lambda^3} + {2 \over \lambda^5}
+ O({1 \over \lambda^7}).
\end{equation}
Next we calculate $X_1$. First choose a cut-off $\epsilon$ which we
will take to be zero at the end. We must take the limit $\delta
\rightarrow 0$ first. Then $X_1$ reduces to the following two pieces
\begin{equation}
\int_0^\epsilon{dx{2
\over \left(\lambda^2 - 1 \right) } \left[
{(x^4 - 3\delta^2x^2) \over (x^2
+\delta^2)^3}\right]} + O(\epsilon),
\end{equation}
and
\begin{equation}
\int_\epsilon^{\sqrt{\lambda^2 - 4}}{dx{2
\over \left(\lambda^2 - 1 -  x^2 \right) } \left[
{1 \over x^2}  - {2 \over\left(\lambda^2 - 1 -  x^2\right)  }\right]}
+ O(\delta^2).
\end{equation}
The first piece is given explicitly by
\begin{equation}
{-2\epsilon^3 \over (1 - \lambda^2)(\delta^2 + \epsilon^2)^2}.
\end{equation}
The second piece reduces to
\begin{equation}
{2 \over (\lambda^2 - 1)\epsilon} - {2 \over 3\sqrt{\lambda^2 - 4}} +
O(\epsilon).
\end{equation}
We see that after taking the $\delta \rightarrow 0$ limit the
singular term of order $1/\epsilon$ cancels and we are left with a
finite result.
Taylor-expanding in powers of $1/\lambda$ yields
\begin{equation}
X_1 = - {2 \over 3\lambda} - {4 \over 3\lambda^3} - {4 \over
\lambda^5} + O({1 \over \lambda^7}).
\end{equation}
Adding the result to $X_2$ obtained in Eq(B.18), we find
\begin{equation}
Im(I_2 + I_3) \sim {1 \over a\lambda^5}.
\end{equation}
Since $I_1 \sim 1/\lambda^5$ as well, the imaginary part of $I$
decreases like $\lambda^{-5}$.

Then we study the case when $\lambda \sim 1^+$. We set $\lambda - 1 = e$
and obtain the imaginary part as a power series expansion in $e$.
We show that the imaginary part behaves like $e^{-3/2}$. We study the
imaginary part of each of the following integrals
\begin{equation}
I_2 ={ i\pi^2 \over 16a}\int_d^{e}{dr{1 \over r} \left[ {1 \over (B
- i\delta)^3} - {1 \over r}{1 \over (B - i\delta)}\right]},
\end{equation}
and
\begin{equation}
I_3 ={ i\pi^2 \over 16a}\int_d^{e + 2}{dr{1 \over r} \left[ {1 \over (A
- i\delta)^3} + {1 \over r}{1 \over (A - i\delta)}\right]}.
\end{equation}
Here, $d$ is a small number that regulates each of the integrals near
$r=0$.
At the end, after taking $d \rightarrow 0$, the sum of $I_2$ and $I_3$ will
turn out to be finite
independent of $d$.

To evaluate the imaginary part of $I_2$, we do the same change of
variables as before, Eq(B.12), and obtain the same expression as
Eq(B.13) but now with the domain of integration being $0 < x < \sqrt{2(e -
d)+e^2}$. For $I_3$, however, we cannot use the same change of variables
as in Eq.(B.14) within the whole domain of integration since, for $r < 1$,
$r - 1$ is negative. When $r < 1$, we must set
\begin{equation}
1 - r = \sqrt{\lambda^2 - x^2}.
\end{equation}
Then the imaginary part of $I_3$ is given by Eq(B.15) plus an
additional term
\begin{equation}
{\pi^2 \over 16a}\int_{\sqrt{2(e +  d)+e^2}}^{\lambda}{dx{1
\over (\lambda^2 - x^2)^{1 / 2}\left[-(\lambda^2 - x^2)^{1 / 2}+1\right] }
\left[
{1 \over x^2} + {1 \over\left[-(\lambda^2 - x^2)^{1 / 2}+1\right]
} \right]}.
\end{equation}

Combining the three pieces together, one is left with the following integrals
\begin{equation}
X_1 = \int_0^{\sqrt{2(e - d)+e^2}}{dx{2
\over \left(2e + e^2 -  x^2 \right) } \left[
{(x^4 - 3\delta^2x^2) \over (x^2
+\delta^2)^3} - {2 \over \left( 2e +e^2 -  x^2\right)  }{x^2 \over (x^2 +
\delta^2)}\right]},
\end{equation}
\begin{equation}
X_2 = \int_{\sqrt{2(e - d)+e^2}}^{\sqrt{2(e + d)+e^2}}{dx{1
\over (\lambda^2 - x^2)^{1 / 2}\left[(\lambda^2  - x^2)^{1 / 2}+1\right] }
\left[
{1 \over x^2} + {1 \over\left[(\lambda^2 - x^2)^{1 / 2}+1\right]
} \right]},
\end{equation}
and
\begin{equation}
X_3 = \int_{\sqrt{2(e+d)+e^2}}^{1+e} dx {2 \over \left[(1+e)^2 -
x^2\right]^{1/2}(x^2-2e -e^2)} \left[ {1 \over x^2} + {(2 + 2e + e^2 -x^2)
\over (x^2 -2e-e^2)}\right].
\end{equation}

First note that
\begin{equation}
X_2 = \int_{\sqrt{2(e - d)+e^2}}^{\sqrt{2(e + d)+e^2}}dx{1\over x^2} +
O(\sqrt{e}).
\end{equation}
This in turn is of order $d$, and, therefore, it vanishes since we take
$d \rightarrow 0$.
The first integral can be calculated as before. The small number
$\delta$ regulates the integral near the lower limit $x=0$, exactly the
same way as before. We are left with
\begin{equation}
X_1 = -{1 \over \sqrt{e(2+e)}d} + finite \ in \ d.
\end{equation}
Here, the finite piece in $d$ is of order $e^{-3/2}$.
The singular term of order $d^{-1}$ arises from the second piece of the
integrand which diverges like
\begin{equation}
1 \over \left[\sqrt{e(2 + e)} - x\right]^2
\end{equation}
near the upper limit of integration.
Finally, consider the integral $X_3$. This should be dominated by the
singular terms near the lower limit of integration. Near the upper
limit of integration the integrand behaves like $1 / \sqrt{1 + e -x}$,
but the integral converges. Therefore, we can expand in powers of $(x^2
- 2e -e^2)$ and consider only the singular terms. We find
\begin{equation}
X_3 = \int_{\sqrt{2(e+d)+e^2}}^{1+e} dx \left[ {1 \over x^2}+ {2 \over
x^2(x^2-2e -e^2)} + {4
\over (x^2 -2e-e^2)^2}\right] + finite.
\end{equation}
The finite piece is finite both in the $d \rightarrow 0$ and $e
\rightarrow 0$ limits.
We find that
\begin{equation}
X_3 = {1 \over \sqrt{e(2+e)}d} + finite \ in \ d,
\end{equation}
and so the singular term of order $d^{-1}$ cancels.
Again, the piece finite in $d$ is of order $e^{-3/2}$.
After taking the $d \rightarrow 0$ limit, we combine the finite
piece in $X_3$ with the finite piece in $X_1$
and expand in powers of $e$, to find
\begin{equation}
X_1 + X_3 = -{1   \over \sqrt{2}e^{3 / 2}} + {7 \over 4
\sqrt{2e}}+ O(e^0).
\end{equation}
Thus the imaginary part of $I$ behaves like
\begin{equation}
Im(I) \sim - {1 \over a(\lambda - 1)^{3 / 2}}.
\end{equation}
We also note that the imaginary part is negative for $\lambda \sim 1^+$.
This was also the case in the $\lambda \gg 1$ limit. Thus we expect
the imaginary part to be increasing negatively as $\lambda \rightarrow 1$.

Finally, let us analyze the case when $\lambda \ll 1$. In this case
$B^2$ is negative and does not contribute to the imaginary part.
We have to extract the imaginary part from the following expression
\begin{equation}
{ i\pi^2 \over 16a}\int_{1 - \lambda}^{1 +\lambda }{dr{1 \over r}
\left[ {1 \over (A
- i\delta)^3} + {1 \over r}{1 \over (A - i\delta)}\right]}.
\end{equation}
We choose to make the following change of variables first
\begin{equation}
u = r - 1.
\end{equation}
The imaginary part is then given by
\begin{equation}
Im(I) ={\pi^2 \over 16a}\int_{-\lambda}^{\lambda}{du{1 \over u+1} \left[
{(A^3 - 3\delta^2A) \over (A^2
+\delta^2)^3} + {1 \over u+1}{A \over (A^2 + \delta^2)}\right]},
\end{equation}
where $A^2 = \lambda^2 - u^2$.
This in turn can be written as follows
\begin{equation}
{\pi^2 \over 8a}\int_{0}^{ \lambda}{du{1 \over 1- u^2} \left[
{(A^3 - 3\delta^2A) \over (A^2
+\delta^2)^3} + {1 +  u^2 \over 1 -  u^2}{A \over (A^2 + \delta^2)}\right]}.
\end{equation}
Now change variables by setting $A^2 = x^2$ to get
\begin{equation}
{\pi^2 \over 8a}\int_{0}^{1}{dx{1 \over \sqrt{1 -
x^2}\left(1 - \lambda^2 + \lambda^2 x^2\right)} \left[
{(x^4 - 3\delta^2x^2) \over \lambda^2 (x^2
+\delta^2)^3} + {\left(1 + \lambda^2 - \lambda^2 x^2\right)
\over\left(1 - \lambda^2 + \lambda^2x^2\right) }{x^2 \over (x^2 +
\delta^2)}\right]},
\end{equation}
where we have rescaled x with $ 1/\lambda$. The second piece in the
integral becomes
\begin{equation}
{\pi^2 \over 8a}\int_{0}^{1}{dx{1 \over \sqrt{1 -
x^2}}} + O(\lambda^2) = {\pi^3 \over 16a} + O(\lambda^2).
\end{equation}
The first piece can be integrated using the same method as
before. Choose a small cut-off
$\epsilon$ and write the integral in terms of
\begin{equation}
L_1 = {\pi^2 \over 8a}\int_{0}^{\epsilon}{dx{
(x^4 - 3\delta^2x^2) \over \lambda^2 (1 - \lambda^2) (x^2
+\delta^2)^3}} + O(\epsilon),
\end{equation}
and
\begin{equation}
L_2 ={\pi^2 \over 8a}\int_{\epsilon}^1{dx{1 \over\lambda^2 x^2 \sqrt{1 -
x^2}\left(1 - \lambda^2 + \lambda^2x^2\right)}} + O(\delta^2).
\end{equation}
Next we observe that
\begin{equation}
L_2 ={\pi^2 \over 8a}\int_{\epsilon}^1{dx{1 \over\lambda^2 (1 - \lambda^2)
x^2 \sqrt{1 -
x^2}}} - {\pi^2 \over 8a}\int_0^1 dx {1 \over \sqrt{1-x^2}}  + O(\lambda^2).
\end{equation}
Evaluating $L_1$ and taking $\delta \rightarrow 0$, we are left with
\begin{equation}
L_1 = -{\pi^2 \over 8a \lambda^2 (1 - \lambda^2) \epsilon}.
\end{equation}
Similarly, after taking $\epsilon \rightarrow 0$, $L_2$ reduces to
\begin{equation}
L_2 = {\pi^2 \over 8a \lambda^2 (1 - \lambda^2) \epsilon} - {\pi^3
\over 16a} + O(\lambda^2).
\end{equation}
Adding the three pieces together, we see that the imaginary part
tends to zero when $\lambda \ll 1$ like $\lambda^2$.

%
%

%
%
%
\setcounter{equation}{0}
%

%
%
%
%
%
\end{document}